\documentclass[twocolumn,amssymb,fleqn, twocolumns]{revtex4-2} 
\usepackage{epsfig,amssymb,amsmath,graphicx,hyperref  }  
\usepackage{color}

\usepackage{graphicx}
\usepackage[caption=false]{subfig}
\usepackage{multirow} 
\usepackage{tabularx}

\newcommand{\be} {\begin{eqnarray}}
\newcommand{\ee} {\end{eqnarray} }

\newcommand{\f} {\frac }

\begin{document}

\title{Geometry and dynamics of spring networks of spherical topology}
\author{Zhenwei Yao}
\email{zyao@sjtu.edu.cn}
\affiliation{School of Physics and Astronomy, and Institute of Natural
Sciences, Shanghai Jiao Tong University, Shanghai 200240, China}
\begin{abstract} 
The spring network model constitutes the backbone in the representations of a
  host of physical systems. In this work, we report the disturbance-driven
  microscopic dynamics of an isolated, closed spring network of spherical topology in
  mechanical equilibrium. The system permits self-intersection. We first show
  the lowest-energy configurations of the closed spring networks as packings of
  regular triangles. The dynamics of the disturbed spring network is analyzed
  from the multiple perspectives of energetics, structural instability, and
  speed distribution.  We reveal the crumpling transition of strongly disturbed
  spring networks and the rapid convergence of the speed distribution toward
  the Maxwell-Boltzmann distribution. This work demonstrates the rich physics
  arising from the interplay of flexibility and dynamics. The results may yield
  insights into the shape fluctuation and structural instability of deformable
  membranes from the dynamical perspective. 
\end{abstract}

\maketitle

\section{Introduction}

The spring network model consists of point particles connected by linear springs of
given stiffness and rest length, and it constitutes the backbone in the
representations of a host of physical systems, including
polymers~\cite{de1979scaling,rubinstein2003polymer}, crystal
lattices~\cite{dong2022exact,liu2025general}, and flexible crystalline
membranes~\cite{Safran2003,Nelson2004c,audoly2010elasticity,vernizzi2011platonic,katsnelson2013graphene}. 
Especially, the 2D spring network model yields insights into the fundamental
physics of elastic membranes from mechanical
deformations~\cite{PhysRevA.38.1005,Lipowsky1995,yong2013elastic,Vella2019} to thermal
fluctuations~\cite{Safran2003,Nelson2004c,kosmrlj2017statistical}. An elastic
membrane refers to a quasi-2D surface of zero or small bending
stiffness. 
The stiffness of the springs is
related to the Young's modulus of the crystalline
membrane~\cite{PhysRevA.38.1005,paulose2012fluctuating}. Furthermore,
disclinations can be naturally embedded into a 2D triangular lattice of linear
springs; an $n$-fold disclination as a fundamental topological defect in
triangular lattice refers to a particle of coordination number $n$ that is
deviated from six~\cite{chaikin2000principles,nelson2002defects}. As such, the
2D spring network model also serves as an ideal platform to explore the physics
of topological defects in crystalline
membranes~\cite{cottrell1965dislocations,Nelson1987,hoffmann2022theory,bhatt2022various}.
For example, a combination of numerical experiments and continuum elasticity
theory revealed the buckling transition caused by the defect-shape
coupling~\cite{PhysRevA.38.1005}, which has a strong connection to the
conformation of virus shells and the organization of particles on flexible
interfaces in
general~\cite{lidmar2003virus,vernizzi2011platonic,bowick2013pathways,funkhouser2013topological,li2018large}.

The mechanical instabilities of crystalline membranes under external pressure,
which are of both scientific and engineering significance, have also been
investigated in the framework of the spring network
model~\cite{paulose2012fluctuating,wan2015effects}. Furthermore, shape
fluctuations of tethered elastic membranes have been systematically
investigated~\cite{Nelson2004c,paulose2012fluctuating,paoluzzi2016shape,chen2022spontaneous}.
An important observation is that the tethered membrane under thermal agitation
exhibits crumpling
transition~\cite{Nelson1987,kantor1987phase,kantor1987crumpling}. Both
renormalization group calculation and Monte Carlo simulations showed that the
energetically unfavored crumpled phase becomes entropically favored at
sufficiently high temperature~\cite{kantor1987phase,kantor1987crumpling}.
Recently, the crumpling phenomenon was examined in active tethered membranes by
introducing active fluctuations into the system; the resulting phase behavior is
overall consistent with that observed for passive
membranes~\cite{gandikota2023crumpling,gandikota2024spontaneous}.

The revealed thermally agitated crumpling of tethered surfaces, which has been
extensively analyzed from the perspective of statistical mechanics, inspired us
to explore the microscopic dynamics in a disturbed spring network of spherical
topology. While the physics of a real elastic membrane that is usually immersed
in an aqueous environment is
complicated~\cite{pozrikidis2001effect,barthes2016motion}, focusing on the
isolated spring network of spherical topology that constitutes the skeleton of
a tethered closed membrane allows us to highlight the disturbance dynamics at
the atomic level. Elucidating the dynamical response of the closed
spring network provides a unique perspective on the fundamental questions
related to the onset of vesicle instability for
functionality~\cite{Pandey2014,xia2010solvent-driven,su2016vesicle} and the
realization of efficient large
deformations~\cite{10942685,041928,forterre2005how,skotheim2005physical}.

In this work, we focus on the spring network of spherical topology. The point
particles and linear springs composing the network are organized into either a
regular triangular lattice or a random amorphous
lattice~\cite{caspar1962physical}. The dynamics is introduced by imposing a
random disturbance to the lowest-energy state. Our dynamics permits
self-intersection. The goal of this work is to explore the disturbance-driven
deterministic microscopic dynamics of both regular and amorphous spring
networks, focusing on the underlying dynamical regularity and structural
instability. The dynamical evolution of the disturbed spring network is obtained
by numerically integrating the equations of motion by the standard Verlet
integration method~\cite{rapaport2004art}. The numerical approach allows us to
track the trajectory of each particle.

We first show the lowest-energy shapes of both regular and amorphous spring
networks as packings of regular elementary triangles. The spring network
exhibits ultra-softness and geometric rigidity, as indicated by the low-energy
excitations of ripples and the lack of floppy modes. The dynamical response of
the spring network upon disturbance is systematically analyzed from the
perspectives of frequency spectra, morphological transformation and speed
distribution among the particles. We reveal the dynamical regularity in the
fluctuating kinetic energy curves by spectral and statistical analysis.
In particular, we highlight the observed crumpling transition in both amorphous and
regular networks under strong disturbance, and systematically discuss the
dynamics of the crumpling transition, including the critical condition and the
dependence of the collapse time on relevant parameters. Compared with an amorphous
network, a regular network withstands a significantly stronger disturbance. We
also discuss the statistical consequence of the disturbance dynamics and show
the thermalization of the system as characterized by the rapid convergence of
the speed distribution over the disturbed networks toward the Maxwell-Boltzmann
distribution.

\section{Model and Method}

\begin{figure}[t]  
\centering 
\includegraphics[width=3.3in]{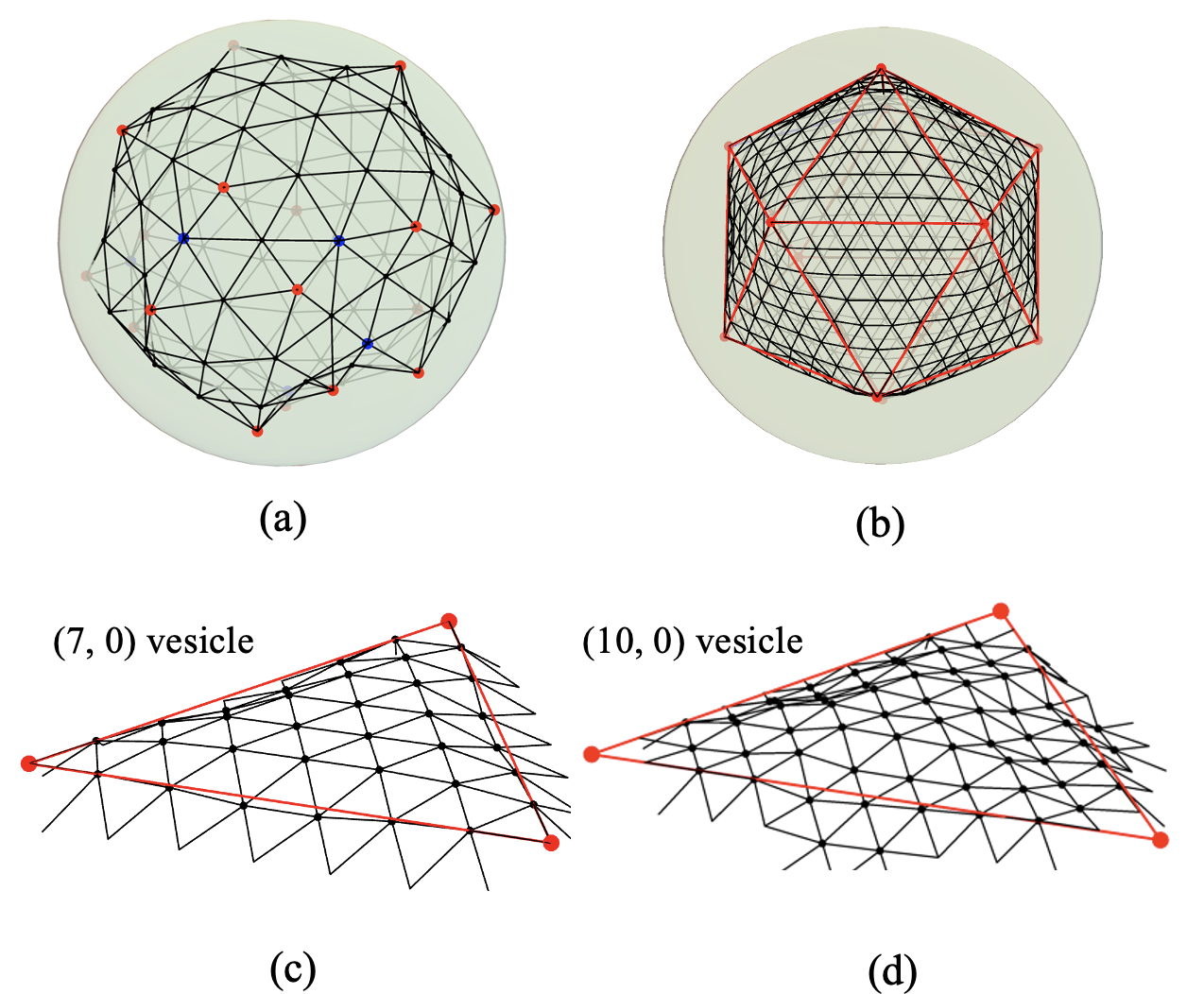}
  \caption{Lowest-energy shapes of amorphous and regular spring networks consist of
  regular triangles. (a) The amorphous network is constructed of randomly
  distributed particles on the sphere.  The total number of particles $V=100$.
  (b) The regular network of $p=7$, and $q=0$. The icosahedral frame of red
  lines is constructed by connecting adjacent disclinations.  Here, for visual
  convenience, a semitransparent plane is inserted in the middle of the spring
  network in (a) and (b); the bonds behind the plane appear semitransparent.
  (c) and (d) Out-of-plane deformations in the lowest-energy shapes of $(7,0)$
  and $(10, 0)$ configurations.  The reference triangles are constructed by
  connecting three adjacent disclinations over the regular spring networks. }
  \label{shape}
\end{figure}

The spring network system of fixed connectivity is constructed from a collection
of point particles of mass $m$ and linear springs of rest length $\ell_0$ and
stiffness $k_0$. For the sake of simplicity, all the springs have the
identical rest length and stiffness in our model. Note that the spring network
model can be extended to explore the heterogeneous phenomena by specifying
spatially varying rest length and stiffness of the springs. Both regular and
amorphous spring networks are investigated in this work. Regular networks are
constructed by the Caspar-Klug scheme~\cite{caspar1962physical}. Specifically,
on a 2D triangular lattice, we select a vector $\vec{G}=p\vec{a} + q \vec{b}$
connecting two points in the lattice, where $\vec{a}$ and $\vec{b}$ are the
elementary lattice vectors that make an angle of $\pi/3$. Then we cut a
regular triangle whose one side coincides with the generating vector $\vec{G}$. By
sticking twenty such identical triangles together side by side in 3D space, we
obtain an icosahedron. The resulting regular network is characterized by the
$(p, q)$ pair, whose value reflects distinct symmetries of the
regular network; see Fig.~\ref{shape}(b). The number of particles is $V=10T +
2$, where the triangulation number $T = p^2 + q^2 + pq$.

Amorphous spring networks are constructed from a collection of point particles
that are initially randomly distributed on a sphere. The random particle
configuration is subsequently relaxed under the Lennard-Jones (L-J) potential
toward a quasi-uniform distribution~\cite{lennard1924ii}. To this end, the
equilibrium distance of the L-J potential is set to be the mean distance of
adjacent particles~\cite{paulose2012fluctuating}. Linear springs between
adjacent particles are then introduced according to the connectivity established
by the Delaunay triangulation on the sphere~\cite{nelson2002defects}. By
removing the geometric constraint of the sphere, the particle-spring network is
relaxed in 3D space toward the lowest-energy state by the steepest descent
method; see Fig.~\ref{shape}(a). The relaxation process is terminated when the
magnitude of the spring deformation corresponding to the maximum force (denoted
as $\delta \ell$) is less than a preset critical value $\delta \ell_c$; $\delta
\ell_c /\ell_0$ is less than $10^{-5}$ in simulations. 

In this work, the length, mass, and time are measured in units of $\ell_0$
(the rest length of spring), $m$ (the mass of a particle), and $\tau_0=
\sqrt{m/k_0}$.

The dynamics is introduced by imposing a random disturbance on the lowest-energy
configuration of the spring network. Specifically, the initial velocity on each
particle is $\vec{v}_{ini} = \Gamma v_0 \vec{\xi}$. $v_0$ is the characteristic
speed.  $v_0=\ell_0/\tau_0=1$. $\vec{\xi}$ is a random vector in 3D space, and
the three independent components conform to the uniform distribution in the
range of $[-1, 1]$. $\Gamma$ is a dimensionless control parameter. Note that the
initial random disturbances on the particles are spatially uncorrelated. We also
specify the initial velocity by the Gaussian distribution, and show the rapid
relaxation of the initial speed distribution within about one characteristic
time. As such, the information about the initial speed distribution is lost in
subsequent dynamical events. More information is provided in the Appendix.

The evolution of the disturbed spring network conforms to the Hamiltonian
dynamics: 
\begin{eqnarray}
  H = \sum_{i\in V} \frac{\vec{p}_i^2}{2m} + \sum_{\alpha \in E}
  \frac{1}{2}k_0(\ell_{\alpha}-\ell_0)^2,
\end{eqnarray}
where the summation is over all the particles and bonds (springs). 
$\vec{p}_i$ is the momentum of particle $i$. $\ell_{\alpha}$ is the length
of spring $\alpha$. We numerically solve for the trajectories of particles by
Verlet integration. The time step $h=10^{-4}$, under which the total energy is
well conserved. Specifically, for the typical spring network system consisting of about
1000 particles, the relative variation of the total energy is at the order
$10^{-6}\%$ during ten million simulation steps.

According to Euler's theorem, topological defects are inevitable in a
triangulated surface of spherical
topology~\cite{Chaikin2000a,bowick2002crystalline}. The fundamental topological
defects in triangular lattice are $n$-fold disclinations. An
$n$-fold disclination refers to a particle of coordination number $n$, and it
carries topological charge $q=(6-n)\pi/3$. By Euler's relation, over a
triangulated surface of spherical topology
\begin{eqnarray}
  \sum_{i} q_i = 4\pi,
\end{eqnarray}
where $q_i$ is the topological charge of particle $i$. Over a regular spring network of
icosahedral symmetry, the twelve five-fold disclinations are located at the
vertices of the icosahedron.

Here, we briefly discuss the connection between the spring network model and
real deformable membranes. The spring network of spherical topology constitutes
the skeleton of a tethered closed membrane.  To fully characterize the behaviors
of real membranes that are usually immersed in aqueous environments or even in
electrolyte solutions, several attributes and constraints shall be added to the
spring network model, including bending
rigidity~\cite{PhysRevA.38.1005,Ou-Yang1999}, charge
distribution~\cite{andelman1995electrostatic,netz2001buckling}, proper dynamics
(hydrodynamics, viscoelasticity,
etc.,)~\cite{evans1976membrane,pozrikidis2001effect,barthes2016motion},
membrane inclusions~\cite{Nelson2004c,ravid2024numerical}, and conservation of
volume or surface area~\cite{Lipowsky1995,Safran2003}.  Here, focusing on the
isolated spring network structure allows us to reveal the intrinsic disturbance
dynamics.

\section{Results and discussion}

This section consists of four subsections. In Sec. III A, we identify the
lowest-energy spring network of spherical topology in mechanical equilibrium.
Specifically, we show the lowest-energy shapes of both regular and amorphous
spring networks as packings of regular triangles. In Sec. III B, we discuss the
dynamical regularity underlying the fluctuating energy curves from the
perspective of spectral analysis. In Sec. III C, we reveal the structural
instability of the closed spring network upon strong disturbance, and
systematically analyze the dynamics of the crumpling transition, including the
critical condition and the dependence of the collapse time on relevant
parameters. In Sec. III D, we discuss the rapid convergence of the speed
distribution over the disturbed networks toward the Maxwell-Boltzmann
distribution. We finally briefly discuss the spring network model and possible
future directions.

\subsection{Lowest-energy shapes: packings of regular triangles}


Prior to solving for the lowest-energy shapes, we first regard the closed spring
network of spherical topology as a geometric frame and analyze its geometric
rigidity~\cite{maxwell1864calculation}. A geometric frame consists of lines
connecting a number of points. The frame is geometrically rigid if the
distance between any two points cannot be altered without altering the length of
one or more lines. In 2D space, a frame of $V$ points requires $2V-3$ connecting
lines (constraints) to render it rigid, where $2V$ is the total number of
degrees of freedom for $V$ particles, and $3$ is the sum of 2 translational
degrees of freedom and 1 rotational degree of freedom. As such, a triangle on
the plane possesses geometric rigidity, but a parallelogram does not. We can
further calculate the number of floppy modes for the frame in 2D space as:
$f=2V-3-E$, where $E$ is the total number of lines. The numbers of floppy modes
for a triangle and a parallelogram are $0$ and $1$, respectively.

For the closed spring network of spherical topology composed of
a triangular lattice in 3D space, the number of floppy modes is 
\begin{eqnarray}
  f=3V-6-E.\label{f}
\end{eqnarray}
$3V$ is the number of degrees of freedom for $V$ particles that can move in 3D
space. $E$ is the number of bonds (springs) in the spring network. The global
rotations and translations contribute six degrees of freedom. To evaluate the
value of $f$, we resort to the relations among $V$, $E$, and $F$. $F$ is the
number of faces (triangles) in the spring network.  According to Euler's
relation, for a triangulated surface of spherical topology,
\begin{eqnarray}
  V-E+F =2. \label{euler}
\end{eqnarray}
Since each face contains three bonds, each of which is shared by two adjacent faces,
we have 
\begin{eqnarray}
  F=2E/3. \label{ef}
\end{eqnarray} 
By combining Eqs.(\ref{f})-(\ref{ef}), we have $f=0$. To conclude, there is no
floppy mode in the spring network of spherical topology regardless of the defect
distribution. The absence of floppy mode indicates that it always costs energy
to deform a spring network in the lowest-energy state.

Here, it is of interest to note that according to Eq.~(\ref{f}), removing bonds
(corresponding to the rupture of the vesicle represented by the spring network)
leads to the emergence of floppy modes. From the thermodynamic perspective, the
rupture of the vesicle allows the system to explore more configurations, and it
is thus an entropically favored process, especially for a large vesicle. In
contrast, when the particles are geometrically confined on the sphere, the
number of degrees of freedom for $V$ particles is changed from $3V$ to $2V$.
Eq.(\ref{f}) thus becomes $f=2V-6-E$. Consequently, we have $f=-V<0$, indicating
that the system is overdetermined. It implies that we can remove bonds in the
spring network confined on the sphere without breaking its geometric rigidity,
because the particles are not allowed to leave the surface of the sphere.


Now, we study the lowest-energy shape of the closed spring network as a
polyhedron.  Polyhedral conformation is a frequently encountered self-assembled
structure, such as polyhedral virus capsids, and compact icosahedra made of
boron oxide~\cite{hubert1998icosahedral} and surfactants~\cite{dubois2001self}.
When confined on the sphere, the residual in-plane stress in the closed spring
network cannot be fully eliminated due to the absence of the isometry mapping
from the plane to the sphere~\cite{audoly2010elasticity}. Geometrically, the
lowest-energy spring network of spherical geometry cannot be composed of
identical regular triangles except the special case of the icosahedral
configuration. Now, by removing the geometric constraint and allowing the
particles to move in 3D space, could the polyhedral spring network of spherical
topology be free of in-plane stress? In other words, we inquire if the lengths
of the springs are uniformly equal to the rest length in the lowest-energy shape
of the spring network.  For a regular spring network, the lowest-energy shapes
include the regular polyhedron consisting of regular triangles and associated
polyhedra generated via local reflection transformations. The case of the
amorphous spring network is more complicated, and we shall resort to numerical
approach to determine the lowest-energy configuration.

\begin{figure*}[t]  
\centering 
\includegraphics[width=6.8in]{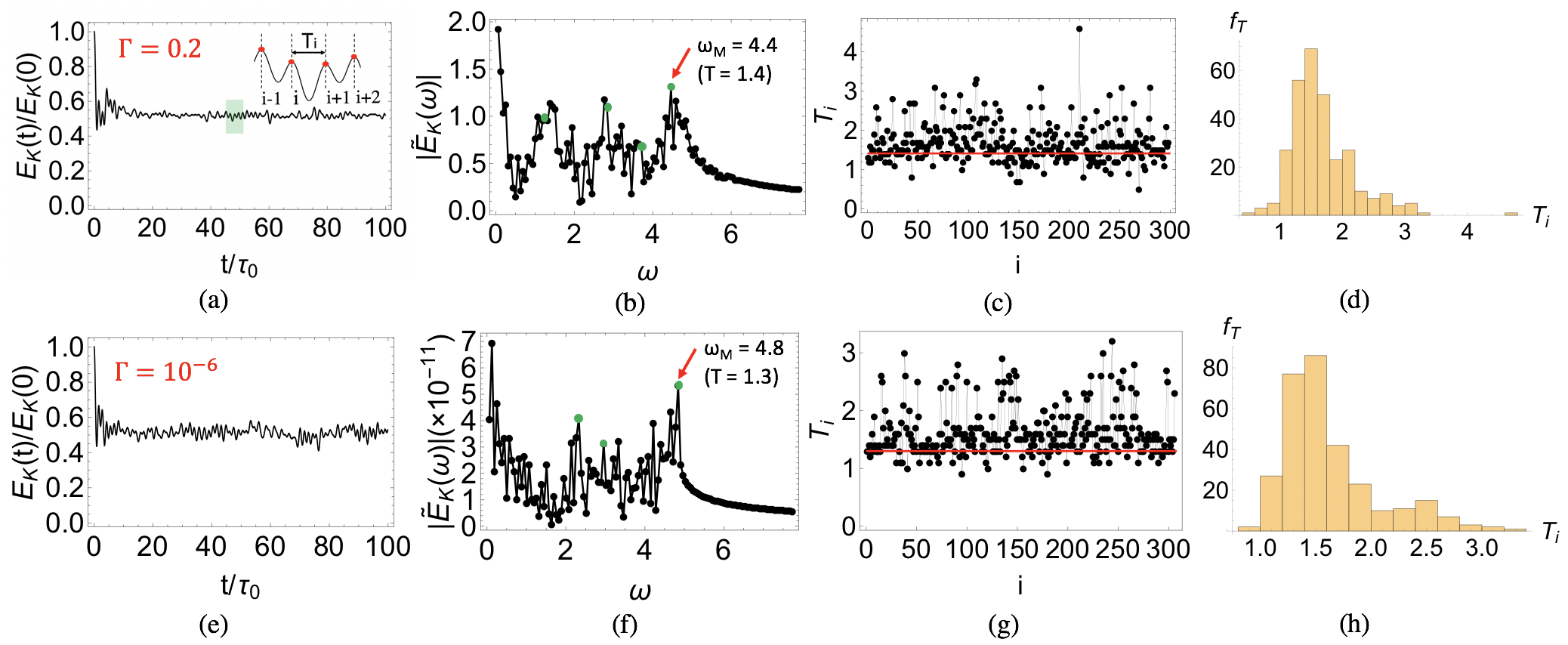}
  \caption{Spectral analysis of the fluctuating kinetic energy curves reveals
  the underlying dominant frequencies.  $\Gamma=0.2$ [(a)-(d)] and
  $\Gamma=10^{-6}$ [(e)-(h)]. In the Fourier transformed kinetic energy curves
  for the time interval $t\in[0, 1000]$ [(b) and (f)], the green dots indicate
  the dominant frequencies during the initial time interval $t\in[0, 200]$.  The
  newly generated frequencies in the evolution of the system are bounded by the
  maximum frequency $\omega_M$ as indicated by the red arrows in (b) and (f).
  The rightmost panels show the statistics of $T_i$, the separation of adjacent
  maxima on the energy curve [see the inset pannel in (a) for the close-up plot
  of the highlighted region]. The red lines in (c) and (g) indicate the values
  of the period corresponding to the maximum frequency $\omega_M$ in (b) and
  (f). $(p, q) = (7, 0)$.  }
  \label{spectrum}
\end{figure*}

We first analyze a group of particles in the spring network: an arbitrary
particle labeled $i$ and its neighboring particles, denoted as $j=1,2,3...z$,
where $z$ is the coordination number and $z \geq 3$. To ensure that the force on
the particle $i$ is zero, there are two possibilities: first, all of the $z$
springs are stress free; second, at least two of the $z$ springs are deformed;
for a buckled particle in particular, at least three springs are deformed. A
particle in the spring network is buckled if a plane can be introduced to separate the
particle and its neighbors. Note that a five-fold disclination is usually
buckled to lower the elastic energy.  Geometric analysis shows that the force on
a buckled particle cannot be zero with only two deformed springs.

The second possibility is excluded in our numerical simulations of amorphous
spring networks constructed from random particle configurations. It is found that all
of the lowest-energy amorphous spring networks consist of regular triangles,
whose sides are equal to the rest length within a small numerical tolerance. A
typical case is presented in Fig.~\ref{shape}(a). For visual convenience, a
semi-transparent plane is inserted in the middle of the spring network; the
bonds behind the plane appear semi-transparent.  The red and blue dots represent
five- and seven-fold disclinations. The buckled five-fold disclinations are
compatible with the intrinsic positive Gaussian
curvature~\cite{bobenko2008discrete}. The entire polyhedron is composed of
regular triangles; the maximum deviation of the spring length from the rest
length is as small as $10^{-10}$.  These results show that once the geometric
constraint of the sphere is removed, the spring network is mechanically relaxed
to the stress-free state.

For regular spring networks, simulations reveal the flexibility feature of the spring network.
In Fig.~\ref{shape}(b), we show the lowest-energy shape of a $(7, 0)$
configuration at the termination condition of $\delta \ell_c/\ell_0 = 10^{-7}$.
For visual convenience, adjacent disclinations are connected to form an
icosahedron.  Closer examination of the shape reveals an appreciable deviation
from the regular icosahedron, as shown in the close-up plot in
Fig.~\ref{shape}(c).  Quantitative measurement shows that the maximum deviation
from the flat triangle spanned by the red lines (the reference triangle) is as
large as $10\%$ of the rest length of the spring. In contrast, the maximum
deviation of the spring length from the rest length is as small as about
$10^{-6}\ell_0$. Is the deviation of the lowest-energy shape from the regular
polyhedron caused by insufficient energy relaxation? To address this question,
we employ a much stricter termination condition of $\delta \ell_c/\ell_0 =
10^{-10}$.  In the resulting lowest-energy shape, the maximum deviation from the
reference triangle is still as large as $10\%$ of the rest length of the spring.
Out-of-plane deformations are also found in lowest-energy regular spring
networks of nonzero $q$ under the same termination condition.

For an even larger regular spring network, we find ripples in the lowest-energy
state.  In Fig.~\ref{shape}(d), we show the local close-up plot of a $(10, 0)$
configuration in the lowest-energy state. The rippled spring network consists of
regular triangles; the maximum deformation of the springs is as small as
$10^{-10}\ell_0$. Note that for the elastic membrane under thermal agitation,
the effect of the resulting shape fluctuations is to
enhance the effective bending rigidity of the membrane~\cite{Nelson2004c}. Here,
the emergent ripple structure may represent the dynamical response of a large
spring network to resist crumpling. From the thermodynamic perspective, the
appearance of the ripples is entropically favored.  Comparison of these
lowest-energy shapes with the perfect polyhedra implies the flexibility of the
spring network in the sense that appreciable deformation of the spring network
occurs under a vanishingly small variation of spring length. In other words, a
small in-plane strain could cause large deformation in 3D space. A similar
geometric effect is also observed in the excitation of ripple structures in
suspended graphene sheets~\cite{meyer2007the}.

We proceed to analyze the distribution of the Gaussian curvature over the
lowest-energy shape of the regular spring network. The discrete version of the
Gauss-Bonnet-Chern theorem for the spherical topology
is~\cite{struik88a,bobenko2008discrete}
\begin{eqnarray}
  \int K dA = \sum_{i=1}^{12} K_i + \sum_{i'=13}^{V} K_{i'} = 2\pi \chi,
\end{eqnarray}
where the Euler characteristic $\chi=2$ for the spherical topology.
The summations in the first and second terms are over the twelve five-fold
disclinations, and the remaining $V-12$ particles. The discrete expression for
the Gaussian curvature at particle $p$ is: $K_p = 2\pi-\sum_{j} \theta_j$.
$\theta_j$ are the interior angles of the triangles meeting at point $p$. Note
that $K_p$ is an intrinsic quantity depending only on the angles of each
triangle and not on the precise embedding of the angles into the 3D Euclidean
space. Simulations show that the value of the Gaussian curvature at each
disclination for the shape in Fig.~\ref{shape}(b) is very close to $\pi/3$. The
summation of these values is very close to $4\pi$; the relative deviation is as
small as $1.6\times 10^{-3}\%$. Since the integral of the Gaussian curvature
over the surface of spherical topology is $4\pi$, the Gaussian curvature in the
shape in Fig.~\ref{shape}(b) is highly concentrated on the disclinations.

\subsection{Dynamical regularity underlying the fluctuating kinetic energy curves}

In the dynamical evolution of the spring network upon a random disturbance, the kinetic
energy is subject to fluctuation; the total energy is well conserved.
Figures~\ref{spectrum}(a) and ~\ref{spectrum}(e) show the highly irregular
kinetic energy curves at varying disturbance strength. In this subsection, we
analyze the dynamical regularity underlying the fluctuating energy curves.

We first perform Fourier transformation on the kinetic energy curves for the time
interval $t\in[0, 1000]$, and the results are presented in
Figs.~\ref{spectrum}(b) and \ref{spectrum}(f).  We see that even under extremely
small disturbance ($\Gamma=10^{-6}$), the system is dominated by a myriad of
frequencies. To demonstrate the nonlinearity driven proliferation of frequencies
in time, the dominant frequencies during the initial time interval $t\in[0,
200]$ are indicated by the green dots in Figs.~\ref{spectrum}(b) and
\ref{spectrum}(f). By extending the time interval up to $t=5000$, we find that
the frequencies are bounded within the maximum frequency $\omega_M$ as
indicated in Figs.~\ref{spectrum}(b) and \ref{spectrum}(f); the values for the
corresponding period are also given. The value of the maximum frequency is
insensitive to the disturbance strength $\Gamma$.

We further investigate the effects of the regularity and size of the spring networks on
the frequency spectrum structure by examining both cases of regular
($p=7$ and $q \in [0, 7]$) and amorphous networks with the same number of
particles. The value of $\Gamma$ is varied in the range of $[10^{-6}, 0.2]$.  In
all these cases, the frequency spectra of the kinetic energy curves are
similar to those in Figs.~\ref{spectrum}(b) and \ref{spectrum}(f). The period
corresponding to the maximum frequency $\omega_M$ is in the range of $1.4\pm
0.1$. This observation suggests that the maximum frequency $\omega_M$ originates
from some local motion. For comparison, $\omega_M$ is much larger than the
frequency of the global breathing mode ($\omega_g$). The breathing mode is
created by imposing a uniform outward radial displacement on each particle as
the initial state.  The resulting eigenfrequency is $\omega_g=0.86$ for the case
of $(7, 0)$ configuration.  Note that $\omega_M$ is also larger than the
eigenfrequency of the free-standing system of two identical masses $m$ connected
by a spring of stiffness $k_0$.  The eigenfrequency of the two-body system is
$\omega = \sqrt{2k_0/m}$. The corresponding frequency of the kinetic energy
curve is doubled: $\omega_{E_k}=2\sqrt{2}\approx 2.8$.


\begin{figure*}[t]  
\centering 
\includegraphics[width=6in]{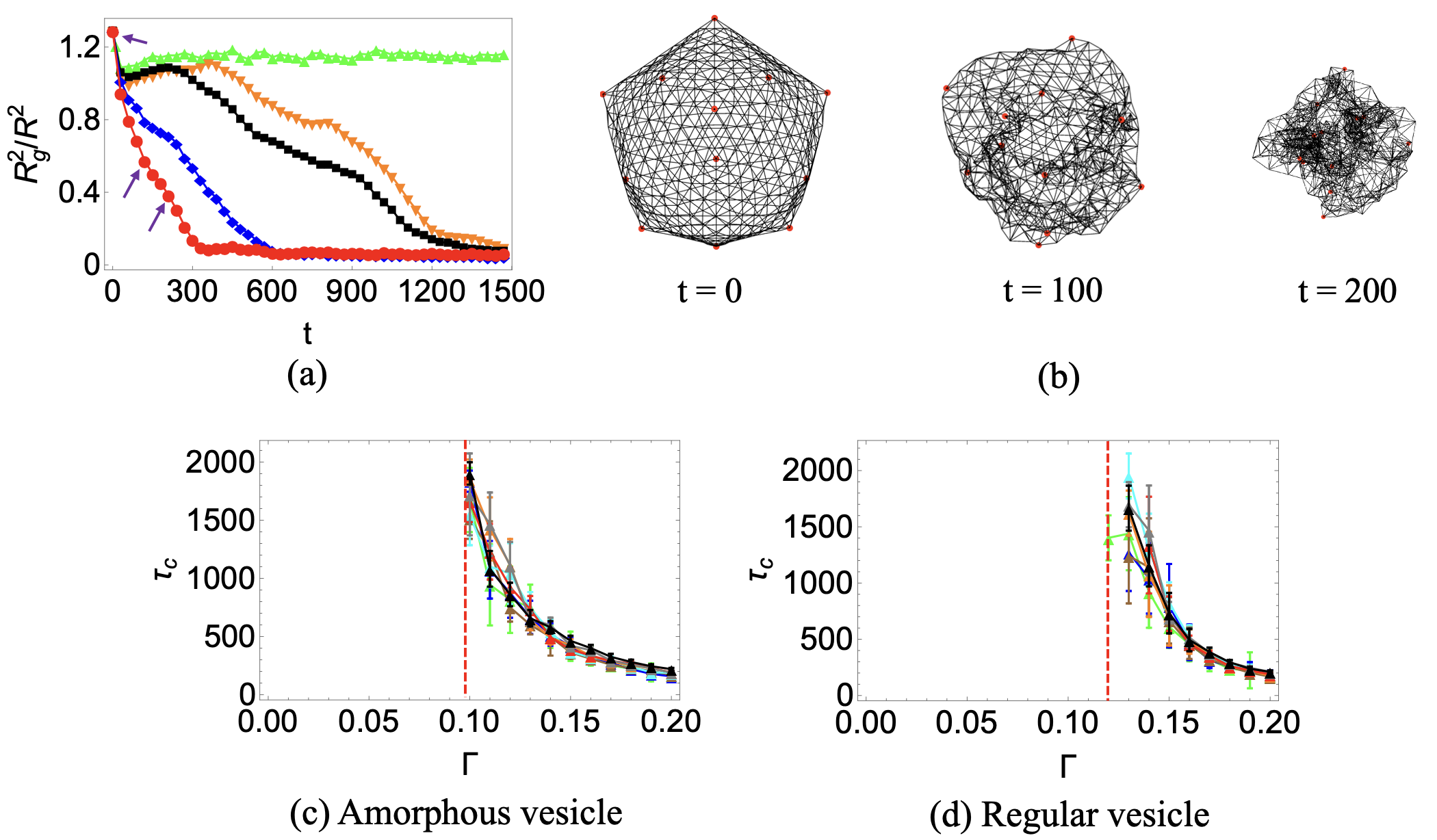}
  \caption{Dynamics of the disturbance-driven crumpling transition of spring
  networks.  (a) Plot of the squared radius of gyration $R_g^2$ (scaled by the
  radius $R$ of the closed spring network) versus time for the $(7,0)$
  configuration.  $\Gamma=0.12$ (green), 0.13 (orange), 0.14 (black), 0.15
  (blue) and 0.16 (red). (b) Instantaneous shapes at the sites indicated by the
  arrows in (a). (c) and (d) show the dependence of the collapse time $\tau_c$
  on $\Gamma$ for both amorphous and regular $(p, q)$ configurations. The curves
  of different colors are for spring networks of varying sizes. In (d), $p=7$,
  $q=0$ (green), 1 (blue), 2 (brown), 3 (cyan), 4 (orange), 5 (red), 6 (gray),
  and 7 (black). The curves of the same color in (c) and (d) are for spring
  networks of identical number of particles.}
  \label{crumpling}
\end{figure*}

To understand the common upper-bounding frequency $\omega_M$ in the frequency
spectra of the kinetic energy curves, we analyze the sequence of the local
maxima on the highly irregular kinetic energy curves. The separation of the
$i$-th and $(i+1)$-th maxima is denoted as $T_i$, as shown in the zoomed-in
inset of the highlighted region in Fig.~\ref{spectrum}(a).  Statistical analysis
of $T_i$ is presented in Figs.~\ref{spectrum}(c)-\ref{spectrum}(d) and
\ref{spectrum}(g)-\ref{spectrum}(h), respectively. The red lines indicate the
values of the period corresponding to $\omega_M$ in Figs.~\ref{spectrum}(b) and
\ref{spectrum}(f).  The concentration of $T_i$ around the red lines is also
shown in the histograms in Fig.~\ref{spectrum}(d) and \ref{spectrum}(h). To
conclude, the maximum frequency $\omega_M$ reflects the dominant rhythm in the
distribution of the time series of $T_i$.  Here, we emphasize that this
dominant frequency is independent of the regularity of the spring network, and
it is almost invariant as the strength of the disturbance is varied by several
orders of magnitude.

\subsection{Dynamics of crumpling transition}

An important observation in the dynamical evolution of the disturbed spring
networks is that both amorphous and regular networks experience crumpling
transition under strong disturbance. Note that an elastic vesicle of spherical
topology is capable of exhibiting interesting morphological transformations like
faceting~\cite{bowick2013pathways}, buckling~\cite{lidmar2003virus}, and
crumpling~\cite{Nelson2004c,wan2015effects}. These featured geometric
transformations have been studied from the perspectives of statistical
mechanics~\cite{Nelson2004c} and the minimization of elastic energy, specifically
under the competition of bending and stretching
energies~\cite{audoly2010elasticity}.

In this subsection, we discuss the crumpling transition from the dynamical
perspective, including the critical condition in terms of the bond strain, and
the dependence of the collapse time on both network size and disturbance
strength.

The morphological transition is characterized by the radius of gyration
$R_g$~\cite{Nelson2004c}:
\begin{eqnarray}
  R_g^2 = \frac{1}{2A^2} \int d^2x \int d^2 x'
  \langle |\vec{r}(x) - \vec{r}(x')|^2 \rangle, \label{Rg2}
\end{eqnarray}
where $A$ is the area of the closed spring network, $\vec{r}(x)$ is the position of the
particle at point $x$ on the surface. In Fig.~\ref{crumpling}(a), we show the
temporal variation of $R_g^2/R^2$ for a regular spring network upon random disturbances
of varying strength. $R$ is the radius of the undisturbed spring network. The values
for $\Gamma$ are $0.12$ (green), 0.13 (orange), 0.14 (black), 0.15 (blue) and
0.16 (red). From Fig.~\ref{crumpling}(a), we see the decline of the $R_g^2/R^2$
curves as $\Gamma$ exceeds $0.12$. The duration of the collapse process, which
is denoted as the collapse time $\tau_c$, is sensitive to the strength of the
disturbance. As $\Gamma$ is increased from $0.13$ to $0.16$, the value for
$\tau_c$ decreases from $1500$ to $300$. On the curves of $\Gamma=0.13$ (orange)
and $\Gamma=0.14$ (black), we notice the appearance of the plateau structure. It
indicates that a spring network may maintain its original morphology for long time up
to a few hundred times of the characteristic time scale $\tau_0$ prior to
crumpling transition. Typical instantaneous morphologies at the moments
indicated by the arrows in Fig.~\ref{crumpling}(a) are presented in
Fig.~\ref{crumpling}(b).

We systematically investigate the crumpling of both amorphous and regular
spring networks of varying size. The dependence of the collapse time $\tau_c$ on the
disturbance strength and network size is summarized in Fig.~\ref{crumpling}(c)
and \ref{crumpling}(d).  In Fig.~\ref{crumpling}(d), we present the cases of
regular spring networks of $p=7$, and $q=0$ (green), 1 (blue), 2 (brown), 3 (cyan), 4
(orange), 5 (red), 6 (gray), and 7 (black). The curves of the same color in
Figs.~\ref{crumpling}(c) and \ref{crumpling}(d) are for spring networks consisting of
identical number of particles. The error bars indicate the standard deviation
based on the statistical analysis of 60 independent simulation runs.

Figures~\ref{crumpling}(c) and \ref{crumpling}(d) show that the collapse time
rapidly decreases with the enhanced strength of the disturbance. The collapse
time is almost unaffected by the system size. The red vertical lines in
Figs.~\ref{crumpling}(c) and \ref{crumpling}(d) indicate the critical values
$\Gamma_c$ for the crumpling transition. $\Gamma_c$ for regular spring networks is
larger by $20\%$ compared with that for amorphous spring networks. Therefore, a
regular spring network is capable of withstanding a significantly stronger
disturbance.


\begin{figure}[t]  
\centering 
\includegraphics[width=2.8in]{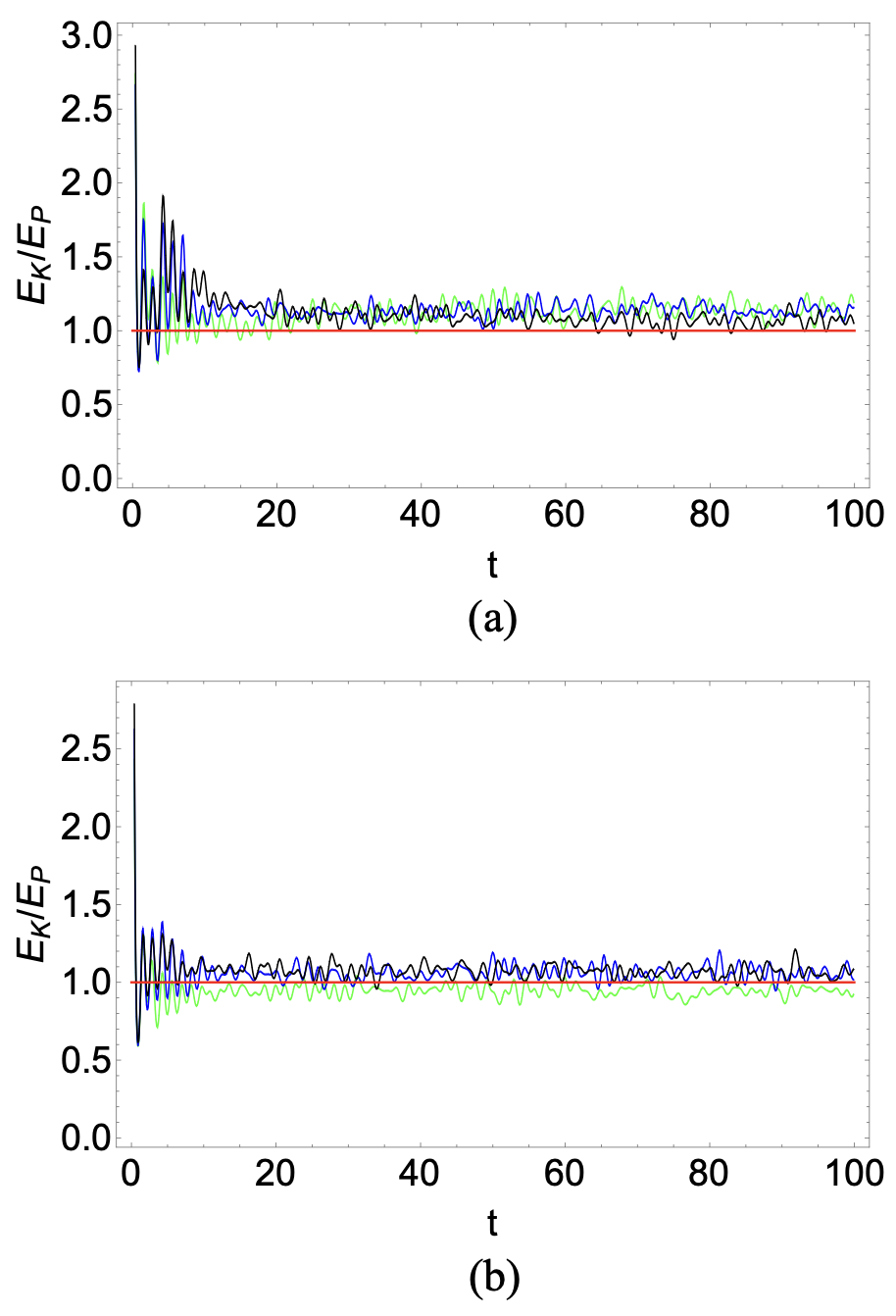}
  \caption{Plot of the ratio of the kinetic and potential energy versus time at
  varying disturbance strength and system size. $\Gamma=0.01$ (green), 0.1
  (blue), 0.2 (black) for the regular $(7, 3)$ configuration (a) and the amorphous
  network (b). The red line indicates the reference position of $E_K/E_P=1$.
  The total number of particles $V=792$ (a) and $1000$ (b).  }
  \label{figEkEp}
\end{figure}

To further understand the occurrence of the crumpling transition, we express the
critical condition in terms of the critical value for the bond strain above
which crumpling transition occurs. In the following, we establish the
relation between the bond strain and the control parameter $\Gamma$ based on
energetic analysis.

In Fig.~\ref{figEkEp}, we show the ratio of kinetic and potential energies
versus time at varying disturbance strength and system size for regular and
amorphous spring networks, respectively. It is found that the ratio $E_K/E_P$
approaches unity in the long run (typically $t > 10$), regardless of the
disturbance strength, the regularity and the size of the spring networks.  The convergence
of the energy ratio to unity is attributed to the presence of a large number of
particles, and it can be understood from the thermodynamic perspective; note
that this ratio would exhibit strong fluctuation in time for a few-particle
system.

Specifically, for a spring network consisting of $V$ particles and $E$ bonds in
thermal equilibrium, according to the theorem of equipartition of energy, the
ratio of the kinetic and potential energy is~\cite{pathria}
\begin{eqnarray} 
  \frac{E_K}{E_P} = \frac{V\times \frac{3}{2}k_B T}{E\times
  \frac{1}{2}k_BT} = \frac{V}{V-2}. \label{therm} 
\end{eqnarray} 
Each particle in 3D space contributes $\frac{3}{2}k_B T$, and each bond
contributes $\frac{1}{2}k_BT$. Equations~(\ref{euler}) and (\ref{ef}) are used
in the last equality. Therefore, the energy ratio approaches unity in the limit
of $V\rightarrow \infty$.

Now, based on the balance of the kinetic and potential energies in equilibrium,
we have
\begin{eqnarray}
  E \times \frac{1}{2}k_0 \langle \Delta \ell^2 \rangle = E_{K}, \label{EkEp}
\end{eqnarray}
where $E$ is the number of bonds. $\Delta \ell = \ell-\ell_0$, where $\ell$ and 
$\ell_0$ are the actual and rest lengths of the spring. $E_{K}$ is the equilibrium
kinetic energy. Since the total energy is conserved and the initial potential
energy is zero, $E_{K} = E_{tot}/2 = E_{K}^{(0)}/2$,
where $E_{K}^{(0)}$ is the initial kinetic energy (i.e., the total energy). 
$E_{K}^{(0)}$ can be computed from the uniform
distribution of the initial velocity:
\begin{eqnarray}
  E_{K}^{(0)} &=& V\times \frac{1}{2}m \sum_{\alpha=x,y,z} \int_{-\Gamma}^{\Gamma}
  v_{\alpha}^2
  p(v_{\alpha}) dv_\alpha \nonumber \\
  &=& \frac{V}{2}m v_0^2 \Gamma^2, \label{Eki}
\end{eqnarray}
where $p(v) = 1/(2\Gamma)$. $v_0$ is the characteristic speed of the system. By
inserting Eq.(\ref{Eki}) into Eq.(\ref{EkEp}), we have
\begin{eqnarray}
  \frac{ \langle \Delta \ell^2 \rangle}{\ell_0^2} &=& \frac{V}{2E} \Gamma^2
  \nonumber \\
  &\approx &   \frac{1}{6} \Gamma^2. \label{ell}
\end{eqnarray}
Equation~(\ref{ell}) establishes the relation between the mean squared bond
strain and the disturbance strength $\Gamma$ for the system in equilibrium.
Note that $\lim_{V\rightarrow \infty} V/E = \lim_{V\rightarrow \infty}
V/(3(V-2)) = 1/3$ according to Eqs.(\ref{euler}) and (\ref{ef}).  $\langle
\Delta \ell^2 \rangle$ is therefore independent of $V$ for $V \gg 1$. This is in
agreement with relevant results in numerical simulations. From the critical
value for $\Gamma_c$, we obtain the critical values: $\sqrt{\langle \Delta
\ell^2 \rangle_c}/\ell_0 \approx 4\%$ and $4.8\%$ for amorphous and regular
spring networks, respectively.

It has been reported that a tethered sheet with preserved connectivity of bonds
and nonzero bending rigidity in thermal equilibrium is capable of exhibiting
transition from flat to crumpled phase without excluded volume interaction
(i.e., the sheet can intersect itself, known as the phantom
surface)~\cite{kantor1986statistical,Nelson2004c}. When the feature of
self-avoidance is included in the model, the tethered sheet exhibits only a flat
phase. Evidences show that flatness of the tethered sheet is an intrinsic
consequence of self-avoidance~\cite{JBOWICK2001255}. These results based on the
statistical mechanics of phantom tethered sheet under thermal agitation shed
light on the disturbance-driven crumpling behavior in our spring network system.
Despite of the distinct agitations, the feature of self-intersection in the
models is crucial for the formation of the crumpled state.


\begin{figure}[t]  
\centering 
\includegraphics[width=3.3in]{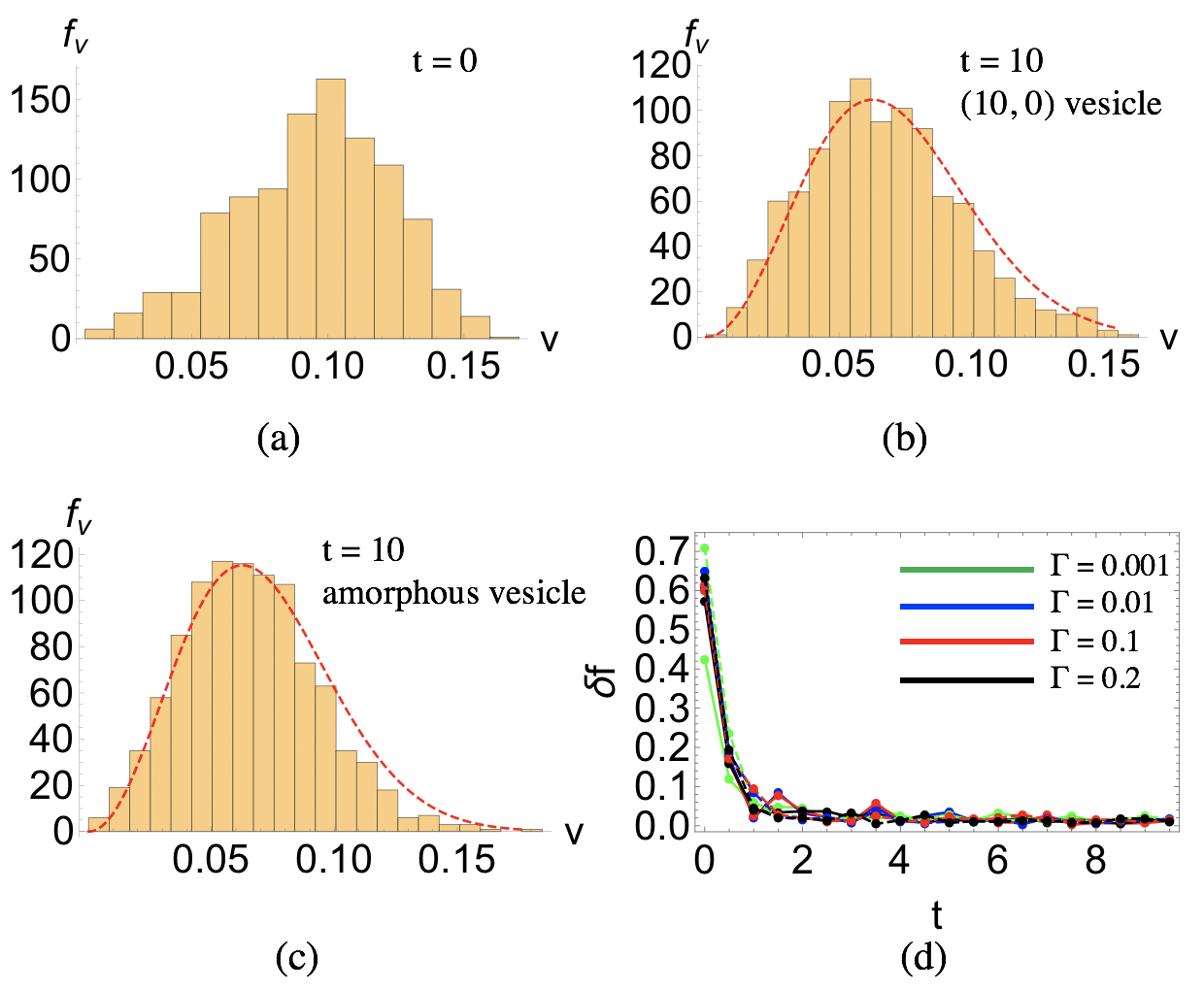}
  \caption{Relaxation dynamics of the speed distribution over the disturbed
  spring networks. (a) Initial speed distribution over a $(10, 0)$ configuration. (b) and (c)
  Convergence of the speed distribution toward the Maxwell-Boltzmann
  distribution, as indicated by the dashed red curves. $\Gamma=0.1$. The total
  number of particles $V=1002$.  (d) Plot of the deviation of the speed
  distribution from the equilibrium Maxwell-Boltzmann distribution versus time.
  The solid and dashed curves are for the regular and amorphous spring networks,
  respectively.  }
  \label{v}
\end{figure}

\subsection{Relaxation dynamics of speed distribution}

In this subsection, we discuss the dynamical evolution of the speed distribution
over the disturbed spring network, and show the rapid thermalization of the
system as a dynamic adaptation to the external disturbance.

In the initial state, the velocity of each particle conforms to uniform
distribution, and the relevant speed distribution is shown in Fig.~\ref{v}(a).
With the evolution of the system, it is found that the peak in the speed
distribution gradually moves leftward toward the low-speed end for both regular
and amorphous spring networks. The speed distributions for both cases evolve toward a
similar profile in long-time simulations, as shown in Figs.~\ref{v}(b) and
\ref{v}(c) for $\Gamma=0.1$. It turns out that the speed distributions can be
well fitted by the Maxwell-Boltzmann distribution: $f_0(v; A,
v_p)=Av^2\exp(-v^2/v_p^2)$, where $v_p$ is the most probable speed and $A$ is
the normalization coefficient. $v_p=15.99$ and $15.96$ in Figs.~\ref{v}(b) and
\ref{v}(c), respectively. Simulations of a series of amorphous spring networks
constructed from independent initial random configurations show that the
convergence of the speed distribution toward the Maxwell-Boltzmann
distribution is unaffected by the specific organization of particles in the
spring network.

We further discuss the dependence of the relaxation time on the disturbance
strength $\Gamma$ and the regularity of the spring network. To quantitatively
characterize the relaxation dynamics, we track the variation of $\delta f$ in
time: 
\begin{eqnarray}
  \delta f(t) = \frac{\int (f_t(v)-f_0(v))^2 dv}{\int f_0(v)^2 dv},
\end{eqnarray}
where $f_t(v)$ is the instantaneous speed distribution at time $t$, and $f_0(v)$
is the Maxwell-Boltzmann distribution. The results are summarized in
Fig.~\ref{v}(d). The solid and dashed curves are for the regular and amorphous
spring networks, respectively. Figure~\ref{v}(d) shows the rapid relaxation of
the speed distribution to the Maxwell-Boltzmann distribution in about one
characteristic time. The relaxation time $\tau_{r}$ thus scales with the
stiffness of spring as $\tau_r \approx \tau_0 \sim k_0^{-1/2}$. The coincidence of the curves
for both regular and amorphous spring networks at varying disturbance strength
$\Gamma$ in the range of $[0.001, 0.2]$ indicates that the relaxation time
$\tau_{r}$ is insensitive to the disturbance strength $\Gamma$ and the
regularity of the spring network.

\

We finally discuss possible extensions of the current work.
Our spring network system represents a highly idealized model without
incorporating the attribute of bending rigidity and the effects of
self-avoidance, gravity, and various impacts from the external environment.
While the ideal spring network model serves as a proper platform to explore the
microscopic dynamical behaviors of interest, it is natural to extend the current
work in the following aspects. First, it is of interest to explore the impact of
bending rigidity on the redistribution of energy over the closed network. In the
static analysis of elasticity, the framework based on the competition of
stretching energy and bending energy is capable of rationalizing a large variety
of buckling phenomena ~\cite{audoly2010elasticity}. An exploration into the
dynamic distribution of energy in the stretching and bending modes is related to
the realization of efficient large deformations. Second, evidence has also
shown the crucial role of self-avoidance in the suppression of the crumpled
phase of the tethered surface under thermal agitation~\cite{Nelson2004c}. The
underlying dynamical scenario is still not clear, which inspires an
investigation of the effect of self-avoidance from the dynamical perspective.
Furthermore, while we focus on the Hamiltonian dynamics of the free-standing
spring network system, it is natural to advance the current work by
incorporating the interaction of the system with the environment. In particular, it is
of interest to examine the dissipative dynamics of the deformable network system
in thermal bath. Under constant thermal agitation that is simultaneously
suppressed by the dissipation, the spring network system provides a proper
platform to explore the Brownian dynamics of a deformable object, which may have
implications for the dynamics of the geometry of elastic vesicles and membranes
in complex fluid environments.

\section{Conclusion}

In summary, we investigated the lowest-energy shapes and the
disturbance-driven dynamics of both regular and amorphous spring networks of
spherical topology. The dynamics of the disturbed spring networks in mechanical
equilibrium has been analyzed from the multiple perspectives of energetics,
structural instability, and speed distribution. We highlight the revealed
dynamical regularity in the fluctuating kinetic energy curves, the crumpling
transition of strongly disturbed spring networks, and the rapid relaxation of
speed distribution. Especially, the feature of self-intersection in our model is
crucial for the formation of the crumpled state. These results advance our
understanding on the shape fluctuation and structural instability of tethered
membranes from the dynamical perspective, and they may have broader implications
considering the extensive applications of the spring network model in
engineering, materials science, and computer graphics.

\section{Acknowledgements}

This work was supported by the National Natural Science Foundation of China
(Grants No. BC4190050).

\begin{figure}[t]  
\centering 
\includegraphics[width=3.45in]{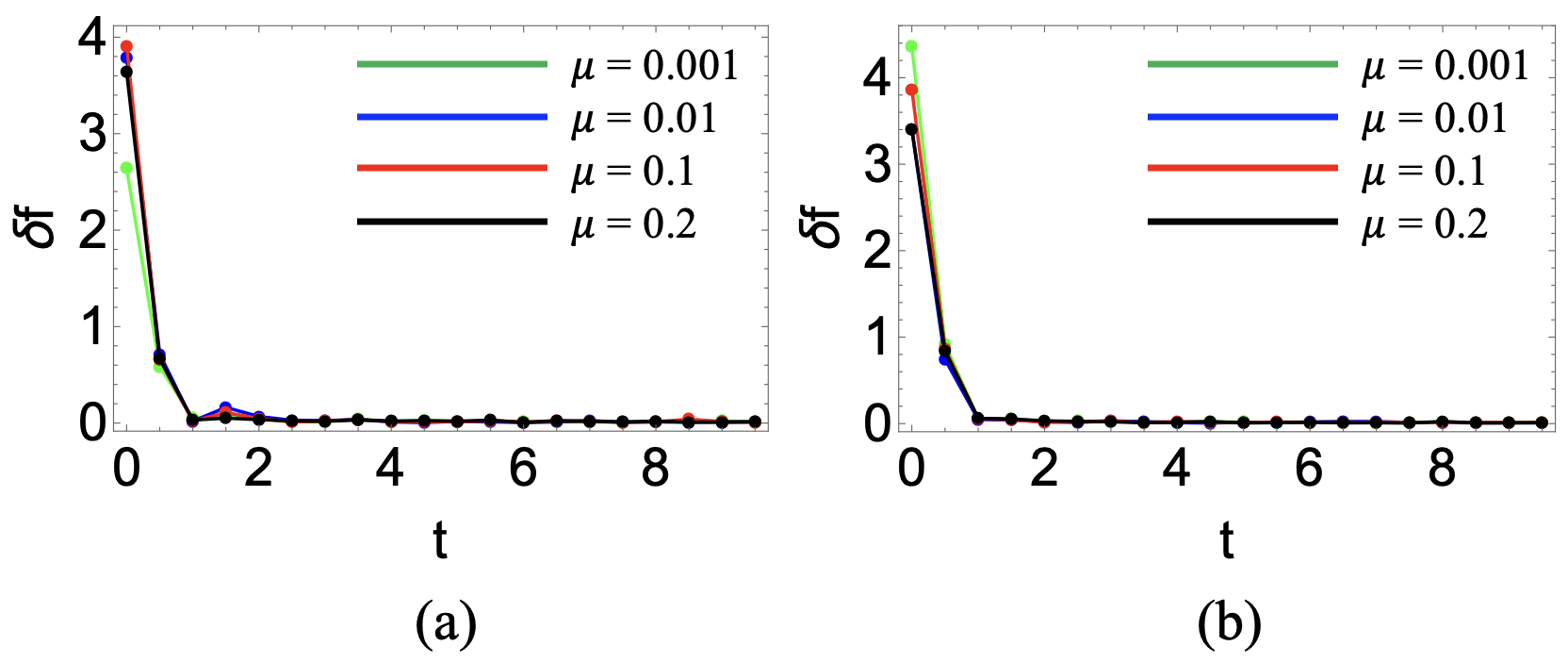}
  \caption{Plot of $\delta f$ (the deviation of the speed distribution from the
  equilibrium Maxwell-Boltzmann distribution) versus time under the initial
  velocity conforming to Gaussian distribution. $\mu$ is the mean value of the
  initial speed. (a) is for the case of regular spring network.  $(p,q) = (10,0)$.
  (b) is for the case of amorphous spring network.  The total number of
  particles $V = 1002$. $\sigma=\mu\times 10\%$. }
  \label{gaussian}
\end{figure}

\section*{Appendix: Initial velocity conforming to Gaussian distribution}

In the main text, the initial velocity on each particle conforms to a uniform
distribution. We also specify the initial velocity using the Gaussian distribution that
preserves the rotational symmetry, and check the subsequent relaxation of speed.
In this Appendix, we present the technical details and the main results on speed
relaxation under the initial velocity conforming to the Gaussian distribution.

The magnitude of the initial velocity conforms to the Gaussian distribution:
\be
p(v) = \f{1}{\sigma \sqrt{2\pi}} e^{-\f{(v-\mu)^2}{2\sigma^2}}, \label{app_1}
\ee
which is characterized by the mean value $\mu$ and the standard deviation
$\sigma$. The strength of the disturbance is characterized by the quantity
$\mu$. The direction of the velocity, as specified by the polar angle $\theta$
and the azimuthal angle $\phi$, is uniformly distributed over the unit sphere.
$\theta \in [0, \pi]$ and $\phi \in [0, 2\pi)$. To create a uniform distribution
of points on the unit sphere, we first introduce the distribution density
function $f(\vec{r})$, where $\vec{r}$ is a unit vector pointing from the center
to the surface of the sphere. The number of points in the solid angle $d\Omega$
is proportional to $f(\vec{r}) d\Omega$, where $d\Omega=\sin\theta d\theta
d\phi$.  For a uniform distribution, $f(\vec{r})$ is a constant;
$f(\vec{r})=1/(4\pi)$ by the normalization condition. To obtain the probability
density function $f(\theta,\phi)$ on the $(\theta, \phi)$-plane that generates a
uniform distribution of points on the sphere, it is required that 
\be
f(\vec{r}) d\Omega = \f{1}{4\pi} d\Omega = f(\theta,\phi) d\theta d\phi,
\ee
from which we have 
\be
f(\theta,\phi) = f_1(\theta) \times f_2(\phi).
\ee
$f_1(\theta)=(1/2)\sin\theta$ and $f_2(\phi)=1/(2\pi)$. Therefore, $\phi$
is a uniformly distributed variable in the range of $[0, 2\pi)$.

To sample the random variable $\theta$ conforming to the distribution
$f_1(\theta)$, we employ the method of inverse transform sampling, which allows
us to sample a general probability distribution using a uniform random number.
Specifically,  we first introduce the cumulative distribution $F(\theta)$ of
$f_1(\theta)$:
\be
F(\theta) = \int_{0}^{\theta} f_1(\theta') d\theta' = \f{1}{2}(1-\cos\theta).
\ee
Note that $F(\theta)$ is a monotonically increasing function of $\theta$ in
$\theta\in [0, \pi]$. The inverse function is denoted as $F^{-1}$.

Now, let $U$ be the uniform random number in $[0,1]$. For any number $x\in
[0,1]$, the probability of $U \leq x$ is equal to $x$, that is, 
\be
Pr(U\leq x) = x. 
\ee
Let $x$ be $F(\theta)$. We have
\be
Pr(U\leq F(\theta)) = F(\theta).
\ee
The following inequality is preserved as $F$ is invertible and monotone:
\be
Pr(F^{-1}(U) \leq \theta) = F(\theta).
\ee
$F(\theta)$ is recognized as the cumulative distribution function for the
random variable $F^{-1}(U)$ as well. As such, $F^{-1}(U)$ follows the same distribution
as the random variable $\theta$. In other words, $\theta$ takes the value of
\be
F^{-1}(u)=\arccos(1-2u),
\ee
where $u$ is a number generated from the uniform
distribution in $[0, 1]$. 

It is of interest to introduce an alternative method of generating uniformly
distributed points on a sphere based on the following geometric
property of the sphere. The area on the sphere between two parallel planes of
equal distance is independent of the position of the planes. Therefore, we can
project the uniformly distributed points on the cylinder to the sphere; the
projection is area preserved.

In Fig.~\ref{gaussian}, we show the variation of $\delta f$ (the deviation of
the speed distribution from the equilibrium Maxwell-Boltzmann distribution) in
time under the initial velocity conforming to Gaussian distribution at varying
$\mu$ for both regular and amorphous spring networks; $\sigma=\mu\times 10\%$.
The number of particles in these two kinds of spring networks are identical.  Note that
in the sampling of the initial speed according to Eq.(\ref{app_1}), no cutoff
is applied to the Gaussian distribution curve. Statistical analysis of the
initial speed in the simulations associated with Fig.~\ref{gaussian} shows that
the maximum deviation from the mean value $\mu$ is within $4\sigma$.

Both Fig.~\ref{gaussian} and Fig.~\ref{v}(d) show that the initial speed
distributions rapidly decay to the equilibrium Maxwell-Boltzmann distribution in
about one characteristic time. We also investigate the cases of larger
dispersion at $\sigma=\mu\times 25\%$ for both regular and amorphous spring
networks, and observe the rapid relaxation of the initial speed distribution in
about one characteristic time, which is the same as in the case of
$\sigma=\mu\times 10\%$ in Fig.~\ref{gaussian}. As such, the information of the
initial speed distribution is lost in subsequent dynamical events, such as the
crumpling transition whose duration is at the order of a hundred characteristic
times.


%

\end{document}